\documentclass[twocolumn,preprintnumbers,amsmath,amssymb]{revtex4} %
\usepackage{epsfig,graphicx}
\usepackage{dcolumn}
\usepackage{bm}

\begin{document}
\draft

\title{The public goods game on homogeneous and heterogeneous networks:
investment strategy according to the pool size}
\author{Zi-Gang Huang,  Zhi-Xi Wu, Jian-Yue Guan, An-Cai Wu, and Ying-Hai Wang\footnote{For correspondence: yhwang@lzu.edu.cn}}
\address{Institute of Theoretical Physics, Lanzhou University, Lanzhou Gansu 730000, China}
\date\today

\begin{abstract}
We propose an extended public goods interaction model to study the
evolution of cooperation in heterogeneous population. The
investors are arranged on the well known scale-free type network,
the Barab\'{a}si-Albert model. Each investor is supposed to
preferentially distribute capital to pools in its portfolio based
on the knowledge of pool sizes. The extent that investors prefer
larger pools is determined by investment strategy denoted by a
tunable parameter $\alpha$, with larger $\alpha$ corresponding to
more preference to larger pools. As comparison, we also study this
interaction model on square lattice, and find that the
heterogeneity contacts favors cooperation. Additionally, the
influence of local topology to the game dynamics under different
$\alpha$ strategies are discussed. It is found that the system
with smaller $\alpha$ strategy can perform comparatively better
than the larger $\alpha$ ones.
\end{abstract}


\maketitle

\section{introduction}

The evolution of cooperation among unrelated individuals is one of
the fundamental problems in biology and social sciences
\cite{vonNeumann,MaynardSmith,HauertScience2002}. A number of
mechanisms have been suggested which are capable of supporting
cooperation in the absence of genetic relatedness. Most notably,
this includes repeated interactions and direct reciprocity
\cite{Trivers,Axelrod}, indirect reciprocity
\cite{Alexander,Nowak3,Nowak4}, punishment
\cite{EFehr,Clutton-Brock}, spatially structured populations
\cite{Nowak1, GSzabo2, Hauert2, Szabo,HauertScience2002}, or
voluntary participation in social interactions
\cite{Szabo,HauertScience2002,Hauert4,Wu1}.

The public goods game (PGG), which attracted much attention from
economists, is a general paradigm to explain cooperative behavior
through group interactions \cite{JHKagel}. In this game, the
defectors who do not contribute, but exploit the public goods,
fare better than the cooperators who pay the cost by contributing.
Thus, the defectors have a higher payoff. If more successful
states spread, cooperation will vanish from the population, and
the public goods along with it. By considering the fact that
who-meets-whom is determined by spatial relationships or
underlying networks, Szab\'{o} and Hauert \emph{et al.} have
recently studied evolutionary PGG in spatially structured
populations bound to regular lattices
\cite{Szabo,Hauert,HauertScience2002} as well as the well-mixed
population \cite{HauertScience2002,Hauert4,Hauert3,Hauert5}. In
these work, the effects of compulsory and voluntary interactions
of agents are also discussed. Several factors such as the
voluntary participation \cite{Hauert}, and small density of
population \cite{Hauert3}, which result in small size of
interaction group, are found to be capable of boosting
cooperation.

However, it has been recognized that regular graphs constitute
rather unrealistic representations of real-world network of
contacts (NoCs), in which local connections (spatial structure)
coexist with long-range connections (or shortcuts). Also,
connections in real-world networks are heterogeneous, in the sense
that different individuals have different numbers of neighbors
with whom they interact. Indeed, these features have been recently
identified as characteristic of a plethora of natural, social and
technological networks
\cite{Dorogotsev2003,Barabasi1999,AlbertBarabasi2002}, which often
exhibit a power-law dependence on their degree distributions. In
addition, it is well accepted that the heterogeneity of network
often plays crucial roles in determining the dynamics
\cite{May2001,Boccaletti,Santos2005}. Therefore it is worth
investigating the public goods interactions on scale-free network.

In this paper, we propose an extended PGG model to study the
evolution of cooperation and investment behavior upon heterogenous
networks. It is known that, in order to reduce risk, investors may
take a portfolio consisting of wide variety of joint enterprise
\cite{portfolio}. In the viewpoint of investors, some enterprise
may be more attractive than others, i.e., there exists
heterogeneity of attractiveness. In our previous work, we have
studied the the effect of heterogeneous influences on the
evolution of cooperation in the prisoner's dilemma game
\cite{Wu2,Wu3,Guan}. Here, in the framework of PGG, we will
consider the effect of heterogeneity of common pools (which we
call pools' ``attractiveness'' $\mathcal{A}$) on investors. In our
model, investors having public goods interactions occupy the
vertices of the underlying network, with the edges denoting
interactions between them. Investors are assumed to be aware of
the sizes of pools, based on which pools' attractiveness can be
estimated. The attractiveness of one given pool also depends on
the investment strategy $\alpha$ of the system, which regulates
the extent of investors' preference to large pools. It will be
shown that, compared with the homogeneous network, the
heterogeneous graphs generated via the mechanisms of growth and
preferential attachment (PA) can distinctively favor cooperation,
where, in addition, smaller $\alpha$ strategy favors cooperation
more than larger $\alpha$ strategy.

\section{our model}
Investors in our model are arranged on certain kind of underlying
network and interact only with their immediate neighbors. A given
investor $j$ acts as an organizer of the common pool $j$ with size
$k_{j}+1$, where there occurs the PGG involving $j$ itself and its
neighbor investors. Here, $k_{j}$ is the degree of $j$. Meanwhile,
investor $j$ will participate in the $k_{j}+1$ pools which we
named the `portfolio' of $j$ including $k_{j}$ pools organized by
its neighbors, and the one pool by itself. The common pool $j$
accumulates capital from all its participant investors, and then
equally shares the profit to them. We assume that investors have
the capacity for learning the sizes of pools in their own
portfolio, which enable them to discriminate pools by estimating
the attractiveness quantitatively. The value of a given pool $i$'s
attractiveness to investors is
$\mathcal{A}_{i}=(k_{i}+1)^{\alpha}$, with the real number
$\alpha$ denoting the investment strategy of the system, which
regulates the extent of investors' preference to larger pools. We
can see that, the larger the $\alpha$ is, the more attractive the
large pools will be. The total amount of capital distributed by a
cooperator (state $s=1$) is normalized to unity, while that of a
defector (state $s=0$) is zero, which implies that defector
withhold its investment to free ride the other investor's
contributions. Investors' capital will be distributed to pools
proportional to the attractiveness. Thus, the amount of capital
investor $j$ distributes to one of its pool $i$ at time $t$ is,
\begin{eqnarray}
D_{ji}(t)=\frac{\mathcal{A}_{i}\cdot s_{j}(t)}{\sum_{l\in
\mathcal{N}(j)}\mathcal{A}_{l}},\label{eqD}
\end{eqnarray}
where $\mathcal{N}(x)$ is the community composed of the nearest
neighbors of $x$ and itself, and $s_{j}(t)$ is the state of
investor $j$ at time $t$. Investors will equally distribute their
capital into pools in portfolio when $\alpha=0$. They invest large
fraction of capital into larger pools for the case of $\alpha>0$,
or into smaller pools for the case of $\alpha<0$. If investor $j$
is a defector [$s_{j}(t)=0$], it distributes nothing to its pools,
and thereby $D_{ji}(t)=0$.

Then, the amount of capital the pool $i$ accumulates at time $t$
can be written as,
\begin{eqnarray}
C_{i}(t)=\sum_{j\in \mathcal{N}(i)}D_{ji}(t) =\sum_{j\in
\mathcal{N}(i)}\frac{\mathcal{A}_{i}\cdot s_{j}(t)}{\sum_{l\in
\mathcal{N}(j)}\mathcal{A}_{l}} \label{eqC}
\end{eqnarray}
We can express Eq. (\ref{eqC}) in terms of the following matrix
equation,
\begin{equation}\label{matrix}
\vec{C}(t)=A\vec{S}(t):=\begin{pmatrix} a_{11} & \dots &
a_{1n}\\
a_{21} & \dots & a_{2n}\\
\hdotsfor{3}\\
a_{n1} & \dots & a_{nn}\\
\end{pmatrix}
\cdot
\begin{pmatrix} s_{1}(t)\\
s_{2}(t)\\
\dots\\
s_{n}(t)
\end{pmatrix}
\end{equation}
where the elements of the matrix $A$ are given by,
\begin{equation}
a_{ij}=
\begin{cases}
{\mathcal{A}_{i}}/{\sum\limits_{l \in \mathcal{N}(j)}\mathcal{A}_{l}}& j \in{\mathcal{N}(i)}\\
0& \text{otherwise}
\end{cases}.
\end{equation}

After the foresaid investing period, the total amount in the pools
multiples by a constant factor $r$, namely, the interest rate on
the common pools. For simplification, we assume that the profit of
each common pool is then equally shared to all participants
irrespective of their individual contribution. Thus the aggregate
payoff of agent $i$ after one generation is,
\begin{eqnarray}
P_{i}(t)=\sum_{j\in \mathcal{N}(i)}\frac{r\cdot C_{j}(t)}{k_{j}+1}
\label{eq07}
\end{eqnarray}
It can be written as $\vec{P}(t)=rBA\vec{S}(t)$, where the
$n\times{n}$ matrix $B$ has element $b_{ij}$ as,
\begin{equation}
b_{ij}=
\begin{cases}
{1}/({k_{j}+1})& j \in{\mathcal{N}(i)}\\
0& \text{otherwise}
\end{cases}.
\end{equation}
Taking into account the unity capital initially distributed by
cooperator, the total returns of investors can be written as
$\vec{R}=\vec{P}-\vec{S}$.
\begin{figure}
\centerline{\resizebox{9.5cm}{!}{\includegraphics{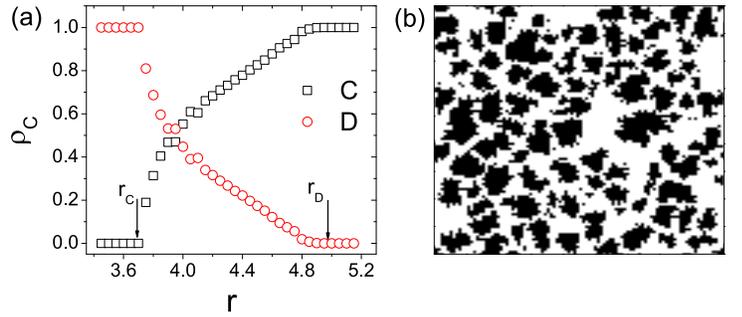}}}
\caption{Average frequencies of cooperators and defectors as a
function of interest rate $r$ (a), and typical snapshot of the
$30\times30$ square lattice with $r=3.8$ (b). Black refers to
cooperators, and white to defectors. The results in (a) are
averaged over $40$ realizations. The $r_{C}$ ($r_{D}$) indicates
the value of $r$ where cooperators (defectors) vanish.}
\label{fig1}
\end{figure}

The return $\vec{R}$ obtained in PGG interactions denotes the
reproductive success, i.e., the probability that one neighbor will
adopt the agent's state. In order to maximize total returns,
investors update their states after each round of game according
to the following rule: Investor $i$ selects one of its neighbor
investor $j$ with equal probability. Given the total returns
($R_{i}$ and $R_{j}$) from the previous round, $i$ adopts neighbor
$j$'s state with probability \cite{GSzabo2,Szabo}:
\begin{eqnarray}
W[s_j\rightarrow{s_i}]=\frac{1}{1+exp[(R_i-R_j+\tau)/\kappa]}
\end{eqnarray}
where $\tau>0$ denotes the cost of state change, and $\kappa$
characterizes the noise introduced to permit irrational choices.
For $\kappa=0$ the neighboring state $s_{j}$ is adopted
deterministically provided the payoff difference exceeds the cost
of state change, i.e., $R_{j}-R_{i}>\tau$. For $\kappa>0$, states
performing worse are also adopted with a certain probability,
e.g., due to imperfect information. Following the previous work
\cite{Szabo}, we simply fix the value of $\kappa$ to be $0.1$.

\section{simulation results}

First, we briefly consider the extended PGG dynamics upon square
lattice with periodic boundary conditions, where the strategy
$\alpha$ does not affect the capital distribution because of the
degree homogeneity. The dynamics starts from the random
arrangement of investors' states either as cooperators or
defectors. In Fig. \ref{fig1}(a), we show the average equilibrium
frequencies of cooperators and defectors as a function of interest
rate $r$. It is expectable that, the curve shows a growth in the
frequency of cooperators with increasing values of $r$. Below the
threshold value $r<r_{C}$ cooperators quickly vanish, whereas for
high $r>r_{D}$ defectors go extinct. For intermediate $r$ the two
states coexist in dynamical equilibrium. The subscript $S$ of
$r_{S}$ refers to the vanishing state. Just as was found in Refs.
\cite{GSzabo2,Nowak1,Szabo}, the snapshot of the dynamics (Fig.
\ref{fig1}(b)) shows that cooperators persist by forming clusters
and thereby minimizing exploitation by defectors.

Let us now consider the evolutionary dynamics of PGG upon the
Barab\'{a}si-Albert (BA) model
\cite{Barabasi1999,AlbertBarabasi2002}. It means that different
investors in this system will have portfolios consisting of
different number of pools. Also, some pools may absorb capital
from many investors, whereas other pools may absorb from much less
investors. In addition, the size of pools exhibits a scale-free
distribution as the degree of the underlying network, which result
in very different values of attractiveness. Here, we want to point
out that, although larger pools have more investors, whether they
can accumulate more capital than the smaller pools still depends
on the investment strategy, preferring the smaller pools
($\alpha<0$), the larger pools ($\alpha>0$), or equally
distributing investment ($\alpha=0$).

\begin{figure}
\centerline{\resizebox{8.5cm}{!}{\includegraphics{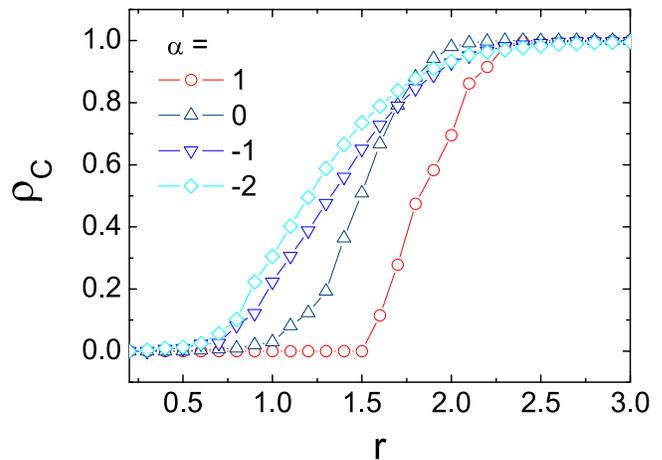}}}
\caption{Average frequency of cooperators as a function of
interest rate $r$ with investment strategy $\alpha=-2$, $-1$, $0$,
and $1$, respectively. Parameters: $\tau=0.1$ and $\kappa=0.1$.}
\label{figPc2}
\end{figure}

Numerical simulations are performed in a system of $N=4000$
investors located on a BA network with average connectivity fixed
as $4$ which is the same as the square lattice. That is, the
network grows from a completely connected network with $m_{0}=5$
vertices, and at every time step a new vertex with $m=2$ edges is
added (the construction of the BA network can refer to Refs.
\cite{Barabasi1999,AlbertBarabasi2002}). Fig. \ref{figPc2} shows
the average equilibrium frequencies of cooperators with different
$\alpha$ strategies on BA networks. One can notice that, no matter
what the strategy $\alpha$ is, the qualitative feature of
frequencies of states remain unchanged for the BA network. Again
three domains are observed: defectors dominate for low $r<r_{C}$,
co-existence for intermediate values of $r$ and homogenous
cooperation for high $r>r_{D}$. However, we find that the
heterogeneous structure distinctly favors cooperators, because the
$r_{D}$ (where defectors vanish) for the BA networks are much
smaller than that for lattice (see Fig. \ref{fig1}). Furthermore,
we notice that the system with smaller $\alpha$ strategy would
behave better than that with larger $\alpha$, which implies that
cooperation are rendered more attractive when investors prefer
smaller pools rather than larger ones.

\section{analysis and discussions}

We can understand the above-mentioned simulation results by the
following analysis. From individual perspective, the unit
investment contributed by one cooperator $i$ will be returned to
the system as $r$ payoff after each round of game. The amount of
payoff returned to $i$ itself can be written as,
\begin{eqnarray}
P_{ii}&=&\sum_{j}\frac{r\cdot D_{ij}(t)}{k_{j}+1}\nonumber\\
&=&r\cdot{\frac{\sum_{j}[1/(k_{j}+1)]\cdot{\mathcal{A}_{j}}\cdot{s_{i}(t)}}{\sum_{j}\mathcal{A}_{j}}}\\
&\equiv& r\cdot{s_{i}(t)}\cdot{L_{\alpha}}. \label{eqPii}
\end{eqnarray}
Here, $j\in{\mathcal{N}(i)}$, and $L_{\alpha}$ denotes the
weighted average of $1/(1+k_{j})$ with $\mathcal{A}_{j}$ as weight
factor. Thus, how much one cooperator can benefit itself from its
own unity investment depends on interest rate $r$, investment
strategy $\alpha$, as well as its \textbf{local topology}
including the degrees of its own and its neighbors.  For a given
investor, the temptation to defect can be measured in terms of
$1.0-P_{ii}$, which is actually independent of other investors'
states. Note that for $r>1/L_{\alpha}$ the social dilemma raised
by the PGG is relaxed in the sense that each unity investment has
a positive net return, and therefore investor can pay off better
when adopting cooperation rather than defection. Here,
$1/L_{\alpha}$ in our extended PGG model in fact corresponds to
the group size in the original PGG model \cite{HauertScience2002}.
For the networks with homogeneous degree $k$, $1/L_{\alpha}$
returns to $k+1$ as the value of group size, while for
heterogeneous network, the values of $1/L_{\alpha}$ are various
for different investors. We give the illustration of $P_{ii}$ on
BA network in the case that all investors adopt cooperation
($\vec{S}=1$). The interest rate $r$ is set as $1.0$ so that
$L_{\alpha}=P_{ii}$.

\begin{figure}
\centerline{\resizebox{10cm}{!}{\includegraphics{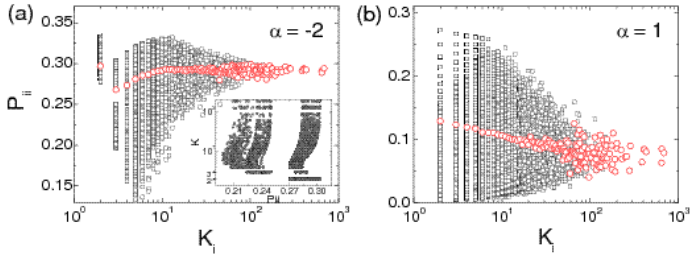}}} \caption{The
$P_{ii}$ of investors as a function of their degree $k_{i}$
(square), and the average $P_{ii}$ over investors with given
degree (dot) with $\alpha=-2$ (a), and $\alpha=1$ (b),
respectively. The network is a BA model with $N=10^{5}$ and
$m=2$.} \label{figPii1}
\end{figure}

In Fig. \ref{figPii1}, we show $P_{ii}$ of investors as a function
of their degree $K_{i}$, with investment strategy $\alpha=-2$ (a)
and $\alpha=1$ (b). It is noteworthy in the figures that investors
with the same degree may have wide range of $P_{ii}$, which
reveals the great diversity of agents' individual local
connection. One can notice from the average values with
$\alpha=-2$ that the $P_{ii}$ of investors with large degree and
smallest degree ($k=2$) is similar, while that of the intermediate
degree investor, especially the $k=3$ investor, is comparatively
small. It is known for BA model that, those agents with degree
$k=m$ are latterly added following degree-PA mechanism, and
thereby more probably to have large degree neighbors. When
investor prefers small pools ($\alpha<0$) the investor $i$ with
degree $k=2$ would distribute most of its investment to pool $i$,
and then gain $1/3$ of the profit, with the largest amount
approximately equal $1/3$ [see Fig. \ref{figPii1}(a)]. In
contrast, agents with $k=3$ still have a neighbor which is younger
than them, thus more likely to be of small degree. The investor
$j$ with $k=3$ will distribute part of its capital $c$ (the amount
between $0.0$ and $1.0$) to its own pool $j$ and then gain $c/4$,
with the remained capital gain less than $(1-c)/3$ profit.
Therefore, generally speaking, the $P_{ii}$ of $k=3$ investor is
smaller than that of $k=2$ investor. In the inset of Fig.
\ref{figPii1}(a), we plot the relations of each $k=3$ investors'
$P_{ii}$ with its $3$ neighbors' degree $k'$. One can notice that,
for the intermediate degree investors, the separated range of
$P_{ii}$ [see Fig. \ref{figPii1}(a)] is closely related to the
$k'=m$ neighbor. Those investors with neighbors' minimum degree
$k'_{min}=2$ would have larger $P_{ii}$, while those with
$k'_{min}\geq3$ are comparatively small ($P_{ii}<1/4$). In
addition, the large degree investors almost deterministically have
$k'_{min}=2$ neighbor pools, which pay them off with more profit
than the large pools. Thus the corresponding ranges of $P_{ii}$
are not separated, and the average values are around $0.3$ for
large degree investors. In addition, when $\alpha=1$ [see Fig.
\ref{figPii1}(b)], investors are likely to distribute most capital
into large pools. The decrease of the average $P_{ii}$ with
agent's degree can be attributed to investor's increasing number
of large pools, which shares the profit to more (other) investors
rather than returning to $i$ itself. From the former analysis, one
gets that local topology of networks has significant impact on PGG
dynamical process.

\begin{figure}
\centerline{\resizebox{9cm}{!}{\includegraphics{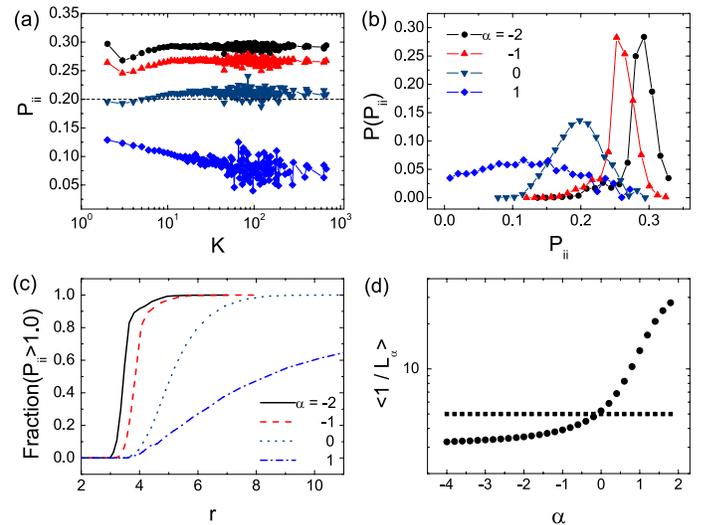}}} \caption{(a)
Average $P_{ii}$ over given degree investors under different
investment strategies. The strategies from top to bottom
respectively are $\alpha=-2$, $-1$, $0$, and $1$. And (b) the
corresponding probability distribution of $P_{ii}$. The interest
rate $r$ is set as $1.0$. (c) The fraction of investors whose
$P_{ii}$ are larger than $1.0$ with the increasing $r$. (d) The
effective group size $\langle1/L_{\alpha}\rangle$ of network
(circle) under different investment strategies $\alpha$, compared
with that of square lattice (square). The network is a BA model
with $N=10^{5}$ and $m=2$.} \label{figPii2}
\end{figure}

In Fig. \ref{figPii2}(a), with respect to the dash line
$P_{ii}=0.2$ which denotes the case of square lattice, we show the
average values of $P_{ii}$ over investors of given degree with
investment strategies $\alpha=-2$, $-1$, $0$, and $1$. The
corresponding probability distributions of $P_{ii}$ are also given
in Fig. \ref{figPii2}(b).  The smaller $\alpha$ strategies are
found to result in comparatively larger values of $P_{ii}$. This
can be easily understood from the form of Eq. (\ref{eqPii})
\cite{Pii}. The fraction of investors who have $P_{ii}>1.0$, with
the increasing of rate $r$ are also plotted in Fig.
\ref{figPii2}(c). We see that, in the system with smaller $\alpha$
strategy, more agents are better off cooperating than defecting
for certain interest rate $r$, no matter what their neighbor
investors do. Furthermore, we can improve our understanding from
the perspective of so-called ``effective group size''
$\langle1/L_{\alpha}\rangle$, which denotes the average impact of
agent's local topology to the game dynamics under certain
investment strategy. The effective group size
$\langle1/L_{\alpha}\rangle$ of BA model with $m=2$ (circle) as a
function of strategy $\alpha$ are shown in Fig. \ref{figPii2}(d).
In the reign of $\alpha<0$, the $\langle1/L_{\alpha}\rangle$ of BA
model is smaller than the group size of the square lattice
(square), while in the reign of $\alpha>0$ the
$\langle1/L_{\alpha}\rangle$ becomes larger. In this point of
view, our result is coincide with that of the former works
\cite{Szabo,Hauert,HauertScience2002,Hauert4,Hauert3,Hauert5},
i.e., smaller group of player favor cooperators in the PGG.

From the illustration of $\vec{S}=1$ case, we know that, agents'
temptation to defect are different because of their different
local topologies. Furthermore, the system with smaller $\alpha$
strategy may render cooperation more attractive for the reason
that cooperator investors can benefit themselves more than those
in systems of larger $\alpha$, which therefore relaxes the social
dilemma better.

\begin{figure}
\centerline{\resizebox{10cm}{!}{\includegraphics{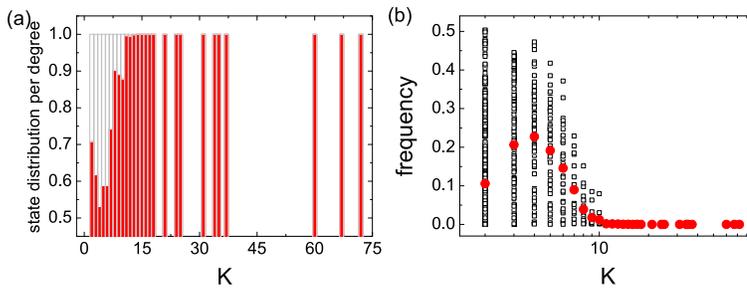}}} \caption{(a)
Distributions of states in BA network. Cooperators are denoted by
red bars, while defectors by white bars, the lengthes of which are
proportional to the relative percentage of the respective state
for each degree $k$. (b) Each agent's frequency of state updating
(open square) during the evolution of the system from the initial
random state to the final equilibrium state within $25000$
generations. And the averages over agents with the same degree
(dot). The simulation takes place on the BA network of size
$N=1000$ and $m=2$, with interest rate $r=1.6$, strategy
$\alpha=0$, which result in the frequency of cooperators
$\rho_{C}=0.684$.} \label{figHis}
\end{figure}

Similar to the former studies of other games
\cite{Santos2005,Santos20061,Santos20062}, the heterogeneity
intrinsic to SF NoCs also contributes to the enhancement of
cooperation, by favoring cooperators to occupy the large degree
agents so as to outperform defectors. The detailed description of
the occupation of vertices with given degree is shown in Fig.
\ref{figHis}(a). One can clearly find that, almost all hubs are
occupied by cooperators, whereas defectors present merely at low
degree vertices. In Fig. \ref{figHis}(b), we plot the state
updating frequency of each agent during the evolution of the
system from the initial state to the final equilibrium state. From
the figures, we know that with respect to the small degree agents,
the hubs always behave as cooperator and rarely change. During the
evolution, when a hub is a defector investor, it can exploit and
may easily invade most of its cooperator neighbors. However, in
doing so, the number of neighbor cooperators will decrease in
subsequent rounds, which in turn acts to reduce the total returns
of such defector hub. Whenever its return becomes comparable to
that of a cooperator neighbor, invasion may occur. On the
contrary, however, once cooperators invade hubs, they will tend to
increase the fraction of cooperator neighbors, in turn maximizing
their own returns. In other words, once invading a hub, a
cooperator becomes so successful that it is very difficult for
defectors to `trike back', as evidenced by the results shown in
Fig. \ref{figHis}.

\section{conclusion}

In summary, considering the heterogeneity of real-world NoCs, we
have proposed an extended public goods interaction model in this
paper. The investor bounded to the underlying network will
distribute capital to pools proportionally to their
attractiveness, which reflects the heterogeneous influence of
pools on investors, with the investment strategy $\alpha$
regulating the value of attractiveness. From the comparative
studies of the game dynamics upon square lattice and BA scale-free
network, we found that heterogeneous structured population
partially resolves the dilemma and improves social welfare. On one
hand, the hub cooperators always remain stable, and spread
cooperation to a larger fraction of the agents. On the other hand,
cooperator investors can pay themselves off with more profit when
taking small $\alpha$ investment strategy, which relaxes the
social dilemma further and enhances the reproductive success of
cooperation.

In addition, the qualitative features of the game dynamics sustain
when the network size $N$ and the parameter $m$ of BA networks are
different. The networks with smaller $m$ are proved to favor
cooperation more, which can be attributed to the corresponding
smaller group size \cite{Hauert,Hauert4}.\\

We thank Dr. Xin-Jian Xu for helpful discussion. This work was
supported by the Fundamental Research Fund for Physics and
Mathematics of Lanzhou University under Grant No. Lzu05008.

\bigskip

\begin{references}
\bibitem{portfolio}
In the real financial system, investors always take a range of
securities to invest. This is because if any particular security
proves to be a bad investment, the impact on such a diversified
portfolio is not as significant as having put all eggs in one
basket.
\bibitem{Pii}
From Eq. (\ref{eqPii}), we know that, for given agent $i$,
$P_{ii}$ would be larger when $\alpha$ is small, because the value
$1/(k+1)$ of the larger $k$ which is small will have more
contribution to $P_{ii}$.
\end{references}
\end{document}